\documentclass[prl,twocolumn,preprintnumbers,superscriptaddress,showpacs,amsmath,amssymb]{revtex4}
\usepackage{graphicx}
\usepackage{dcolumn}
\usepackage{bm}
\usepackage{color}

\begin{document}

\title{Observation of enhanced chiral asymmetries in the inner-shell photoionization \\of uniaxially oriented methyloxirane enantiomers}

\author{M. Tia}\email{tia@atom.uni-frankfurt.de}
\affiliation{Institut f\"{u}r Kernphysik, Universit\"{a}t Frankfurt, Max-von-Laue-Stra{\ss}e 1, 60438 Frankfurt am Main, Germany}

\author{M. Pitzer}
\affiliation{Institut f\"{u}r Kernphysik, Universit\"{a}t Frankfurt, Max-von-Laue-Stra{\ss}e 1, 60438 Frankfurt am Main, Germany}
\affiliation{Institut f\"{u}r Physik und CINSaT, Universit\"{a}t Kassel, Heinrich-Plett-Str. 40, 34132 Kassel, Germany}

\author{G. Kastirke}
\affiliation{Institut f\"{u}r Kernphysik, Universit\"{a}t Frankfurt, Max-von-Laue-Stra{\ss}e 1, 60438 Frankfurt am Main, Germany}

\author{J. Gatzke}
\affiliation{Institut f\"{u}r Kernphysik, Universit\"{a}t Frankfurt, Max-von-Laue-Stra{\ss}e 1, 60438 Frankfurt am Main, Germany}

\author{H.-K. Kim}
\affiliation{Institut f\"{u}r Kernphysik, Universit\"{a}t Frankfurt, Max-von-Laue-Stra{\ss}e 1, 60438 Frankfurt am Main, Germany}

\author{F. Trinter}
\affiliation{Institut f\"{u}r Kernphysik, Universit\"{a}t Frankfurt, Max-von-Laue-Stra{\ss}e 1, 60438 Frankfurt am Main, Germany}

\author{J. Rist}
\affiliation{Institut f\"{u}r Kernphysik, Universit\"{a}t Frankfurt, Max-von-Laue-Stra{\ss}e 1, 60438 Frankfurt am Main, Germany}

\author{A. Hartung}
\affiliation{Institut f\"{u}r Kernphysik, Universit\"{a}t Frankfurt, Max-von-Laue-Stra{\ss}e 1, 60438 Frankfurt am Main, Germany}

\author{D. Trabert}
\affiliation{Institut f\"{u}r Kernphysik, Universit\"{a}t Frankfurt, Max-von-Laue-Stra{\ss}e 1, 60438 Frankfurt am Main, Germany}

\author{J. Siebert}
\affiliation{Institut f\"{u}r Kernphysik, Universit\"{a}t Frankfurt, Max-von-Laue-Stra{\ss}e 1, 60438 Frankfurt am Main, Germany}

\author{K. Henrichs}
\affiliation{Institut f\"{u}r Kernphysik, Universit\"{a}t Frankfurt, Max-von-Laue-Stra{\ss}e 1, 60438 Frankfurt am Main, Germany}

\author{J. Becht}
\affiliation{Institut f\"{u}r Kernphysik, Universit\"{a}t Frankfurt, Max-von-Laue-Stra{\ss}e 1, 60438 Frankfurt am Main, Germany}

\author{S. Zeller}
\affiliation{Institut f\"{u}r Kernphysik, Universit\"{a}t Frankfurt, Max-von-Laue-Stra{\ss}e 1, 60438 Frankfurt am Main, Germany}

\author{H. Gassert}
\affiliation{Institut f\"{u}r Kernphysik, Universit\"{a}t Frankfurt, Max-von-Laue-Stra{\ss}e 1, 60438 Frankfurt am Main, Germany}

\author{F. Wiegandt}
\affiliation{Institut f\"{u}r Kernphysik, Universit\"{a}t Frankfurt, Max-von-Laue-Stra{\ss}e 1, 60438 Frankfurt am Main, Germany}

\author{R. Wallauer}
\affiliation{Institut f\"{u}r Kernphysik, Universit\"{a}t Frankfurt, Max-von-Laue-Stra{\ss}e 1, 60438 Frankfurt am Main, Germany}

\author{A. Kuhlins}
\affiliation{Institut f\"{u}r Kernphysik, Universit\"{a}t Frankfurt, Max-von-Laue-Stra{\ss}e 1, 60438 Frankfurt am Main, Germany}

\author{C. Schober}
\affiliation{Institut f\"{u}r Kernphysik, Universit\"{a}t Frankfurt, Max-von-Laue-Stra{\ss}e 1, 60438 Frankfurt am Main, Germany}

\author{T. Bauer}
\affiliation{Institut f\"{u}r Kernphysik, Universit\"{a}t Frankfurt, Max-von-Laue-Stra{\ss}e 1, 60438 Frankfurt am Main, Germany}

\author{N. Wechselberger}
\affiliation{Institut f\"{u}r Kernphysik, Universit\"{a}t Frankfurt, Max-von-Laue-Stra{\ss}e 1, 60438 Frankfurt am Main, Germany}

\author{P. Burzynski}
\affiliation{Institut f\"{u}r Kernphysik, Universit\"{a}t Frankfurt, Max-von-Laue-Stra{\ss}e 1, 60438 Frankfurt am Main, Germany}

\author{J. Neff}
\affiliation{Institut f\"{u}r Kernphysik, Universit\"{a}t Frankfurt, Max-von-Laue-Stra{\ss}e 1, 60438 Frankfurt am Main, Germany}

\author{M. Weller}
\affiliation{Institut f\"{u}r Kernphysik, Universit\"{a}t Frankfurt, Max-von-Laue-Stra{\ss}e 1, 60438 Frankfurt am Main, Germany}

\author{D. Metz}
\affiliation{Institut f\"{u}r Kernphysik, Universit\"{a}t Frankfurt, Max-von-Laue-Stra{\ss}e 1, 60438 Frankfurt am Main, Germany}

\author{M. Kircher}
\affiliation{Institut f\"{u}r Kernphysik, Universit\"{a}t Frankfurt, Max-von-Laue-Stra{\ss}e 1, 60438 Frankfurt am Main, Germany}

\author{M. Waitz}
\affiliation{Institut f\"{u}r Kernphysik, Universit\"{a}t Frankfurt, Max-von-Laue-Stra{\ss}e 1, 60438 Frankfurt am Main, Germany}

\author{J. B. Williams}
\affiliation{Institut f\"{u}r Kernphysik, Universit\"{a}t Frankfurt, Max-von-Laue-Stra{\ss}e 1, 60438 Frankfurt am Main, Germany}
\affiliation{Department of Physics, University of Nevada, Reno, Nevada 89557, USA}

\author{L. Schmidt}
\affiliation{Institut f\"{u}r Kernphysik, Universit\"{a}t Frankfurt, Max-von-Laue-Stra{\ss}e 1, 60438 Frankfurt am Main, Germany}

\author{A. D. M\"{u}ller}
\affiliation{Institut f\"{u}r Physik und CINSaT, Universit\"{a}t Kassel, Heinrich-Plett-Str. 40, 34132 Kassel, Germany}

\author{A. Knie}
\affiliation{Institut f\"{u}r Physik und CINSaT, Universit\"{a}t Kassel, Heinrich-Plett-Str. 40, 34132 Kassel, Germany}

\author{A. Hans}
\affiliation{Institut f\"{u}r Physik und CINSaT, Universit\"{a}t Kassel, Heinrich-Plett-Str. 40, 34132 Kassel, Germany}

\author{L. Ben Ltaief}
\affiliation{Institut f\"{u}r Physik und CINSaT, Universit\"{a}t Kassel, Heinrich-Plett-Str. 40, 34132 Kassel, Germany}

\author{A. Ehresmann}
\affiliation{Institut f\"{u}r Physik und CINSaT, Universit\"{a}t Kassel, Heinrich-Plett-Str. 40, 34132 Kassel, Germany}

\author{R. Berger}
\affiliation{Theoretical Chemistry, Universit\"{a}t Marburg, Hans-Meerwein-Stra{\ss}e, 35032 Marburg, Germany}

\author{H. Fukuzawa}
\affiliation{Institute of Multidisciplinary Research for Advanced Materials, Tohoku University, Sendai 980-8577, Japan}

\author{K. Ueda}
\affiliation{Institute of Multidisciplinary Research for Advanced Materials, Tohoku University, Sendai 980-8577, Japan}

\author{H. Schmidt-B\"{o}cking}
\affiliation{Institut f\"{u}r Kernphysik, Universit\"{a}t Frankfurt, Max-von-Laue-Stra{\ss}e 1, 60438 Frankfurt am Main, Germany}

\author{R. D\"{o}rner}
\affiliation{Institut f\"{u}r Kernphysik, Universit\"{a}t Frankfurt, Max-von-Laue-Stra{\ss}e 1, 60438 Frankfurt am Main, Germany}

\author{T. Jahnke}
\affiliation{Institut f\"{u}r Kernphysik, Universit\"{a}t Frankfurt, Max-von-Laue-Stra{\ss}e 1, 60438 Frankfurt am Main, Germany}

\author{Ph. V. Demekhin}\email{demekhin@physik.uni-kassel.de}
\affiliation{Institut f\"{u}r Physik und CINSaT, Universit\"{a}t Kassel, Heinrich-Plett-Str. 40, 34132 Kassel, Germany}

\author{M. Sch\"{o}ffler}\email{schoeffler@atom.uni-frankfurt.de}
\affiliation{Institut f\"{u}r Kernphysik, Universit\"{a}t Frankfurt, Max-von-Laue-Stra{\ss}e 1, 60438 Frankfurt am Main, Germany}

\date{\today}

\begin{abstract}
Most large molecules are chiral in their structure: they exist as two enantiomers, which  are mirror images of each other. Whereas the rovibronic sublevels of two enantiomers are almost identical, it turns out that the photoelectric effect is sensitive to the absolute configuration of the ionized enantiomer -- an effect termed Photoelectron Circular Dichroism (PECD). Our comprehensive study demonstrates that the origin of PECD can be found in the molecular frame electron emission pattern connecting PECD to other fundamental photophysical effects as the circular dichroism in angular distributions (CDAD). Accordingly, orienting a chiral molecule in space enhances the PECD by a factor of about 10.

\end{abstract}

%\keywords{Photon interactions with molecules; Auger effect and inner-shell excitation or ionization; Optical activity and dichroism; Chiral media}

\maketitle

\begin{figure}
\includegraphics[scale=0.155]{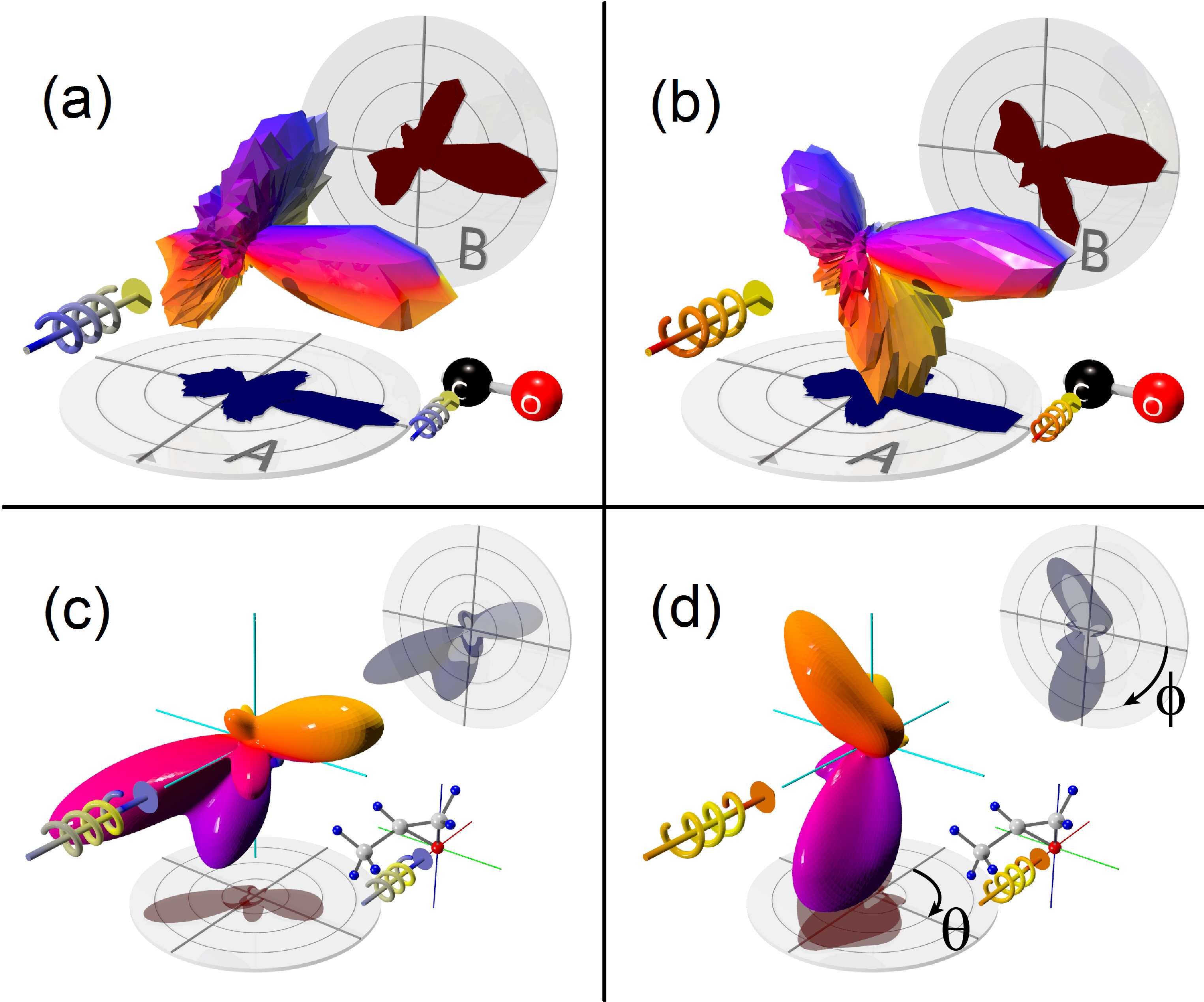}
\caption{(Color online) Three-dimensional molecular frame photoelectron angular distributions. Top: for the C 1s-electrons emitted from CO for left (a) and right (b) circularly polarized light (taken from Ref. \cite{r05}). Bottom: theoretical distributions computed here for the O 1s-ionization of R(+) methyloxirane by left (c) and right (d) circularly-polarized light. The molecules are oriented as depicted in the insets. {Polar and azimuthal angles $\left\{\theta,\phi\right\}$ are indicated in panel (d).} }\label{fig:fig1}
\end{figure}

Photoionization of unpolarised electronic states of an atom is insensitive to the light's helicity: The count rate on an electron detector placed at any particular angle does not change when switching between photons of different circular polarization. In order to make the photoelectron count rate sensitive to the photon helicity, the measurement conditions need to establish a coordinate frame of specific handedness. Two of the three vectors required to define such a coordinate frame are the $k$-vector of the photon (i.e., the photon propagation direction) and the $k$-vector of the photoelectron (i.e., the emission direction). The third vector can be introduced, for instance, by a second photoelectron in the case of photo double ionization, where the observed coincident detection of two electrons depends on the light's helicity \cite{r01,r02}. Alternatively, orienting a linear molecule in space can provide an additional (molecular) axis, which results in a prominent effect that depends on the helicity of the light \cite{r03,r04}. Results from the well-studied example \cite{r04,r05} of the inner-shell ionization of CO are shown in Fig.~\ref{fig:fig1}a and \ref{fig:fig1}b. While the photoelectron angular emission distribution possesses a strong asymmetry within the light's polarization plane (Plane B in Fig.~\ref{fig:fig1}), its forward/backward symmetry (i.e. the symmetry relative to the photon propagation direction in Plane A) remains.

In the case of molecular photoionization, the shape of the emission distribution results from the scattering of the outgoing electron wave by the molecular ion potential. The circularly polarized light additionally imprints the direction of the rotation of its electric field onto that scattered wave. Because switching the helicity of the light is equivalent to a parity inversion, it results in the inversion of the distribution in Fig.~\ref{fig:fig1}a along the vertical axis, yielding the emission pattern shown in Fig.~\ref{fig:fig1}b. The normalized difference of these emission patterns for the two helicities is known as the Circular Dichroism in Angular Distributions (CDAD, \cite{r04}) and since it is symmetric in the forward/backward directions, it depends only on the azimuthal angle $\phi$ around the light propagation \cite{r05}: $CDAD=\frac{I_{+1}(\phi)-I_{-1}(\phi)}{I_{+1}(\phi)+I_{-1}(\phi)}$. Here $I_{+1}$ and $I_{-1}$ correspond to the ionization cross section by left or right circularly polarized light (labeled by positive or negative helicity $\pm1$). The asymmetry of the electron flux induced by circularly polarized photons in the up/down directions has been successfully utilized, e.g., in surface science to stereoscopically image nearest neighbour distances \cite{r06}. For molecules which are \emph{fixed in space}, CDAD values up to 100\% can be observed \cite{r05}.

A forward/backward asymmetry in the photoemission -- even within the electric-dipole approximation \cite{r07} -- can arise due to scattering of the electron wave in the molecular potential whenever the structure of a fixed-in-space molecule breaks that symmetry. The change of this forward/backward asymmetry in the emission distribution that arises when switching the light's helicity is termed Photoelectron Circular Dichroism (PECD). PECD occurs as the molecular structure acts as a gearbox, which translates the rotation of the electric field vector into a change of the forward (or backward) directed electron flux. A mechanical analogue for such machinery is a nut on a thread. The thread transforms the rotation of the nut into forward (or backward) directed motion. Figures~\ref{fig:fig1}c and \ref{fig:fig1}d show the corresponding effect on the molecular level: our calculated electron emission patterns from fixed in space R(+) methyloxirane {(C$_3$H$_6$O)} show dramatic changes upon switching the light helicity and thus substantial PECD. Furthermore, from these figures it becomes intuitively understandable, that PECD is sensitive to subtle changes of the molecular potential (shape and structure) both, static \cite{r08,r09,r10,r11,r12,r13} and dynamic \cite{r14}.

Only for non-racemic mixtures of chiral molecules a PECD is even observable after averaging over all possible molecular orientations \cite{r15,r16}, while for achiral molecules all asymmetries cancel out for a sample of randomly oriented molecules. This is due to the fact that--for achiral molecules--the mirror image of any molecular orientation can be created by a rotation, and by definition the PECD value has the opposite sign for its mirror image configuration. For non-racemic chiral molecules on the contrary, the cancelation can be incomplete, as the mirror situation with the opposite sign of the PECD equals a switch of enantiomers and thus cannot be generated by a rotation. In the last decade, PECD for randomly oriented molecules has been invoked as a powerful chiroptical tool to enable determination of the absolute configuration of chiral molecules \cite{r17,r18,r19}. Recently, the first laser-based PECD measurements \cite{r20,r21,r22,r23} have further demonstrated their potential as analytical applications for characterization of chiral pharmaceuticals. PECD has also been speculated to be one of the symmetry-breaking mechanism at the origin of life's homochirality \cite{r11}.

 While it has already been suggested that the scattering of the photoelectron wave at the molecular potential is at the heart of PECD \cite{r15,r24,r25}, the validity of this intuitive picture has so far not been demonstrated directly, for example, by performing experiments on spatially oriented molecules. Contrarily, to date, PECD in the gas phase has only been studied for randomly oriented molecules. {Accordingly, the observed effect is comparably weak in those studies.} A typical magnitude of the normalized difference \cite{r24}, $PECD=\frac{I_{+1}(\theta)-I_{-1}(\theta)}{I_{+1}(\theta)+I_{-1}(\theta)}$, where $\theta$   is the polar emission angle of the electron with respect to the light propagation, was on the order of a few percent, because integration over all molecular orientations drastically reduces the contrast and thus lowers the PECD values. Note that several definitions of PECD exist in literature, and the present PECD refers to the dichroic parameter $b_1$, which is half of the routinely used $2b_1$.

Our calculations shown in Figs.~\ref{fig:fig1}c and \ref{fig:fig1}d demonstrate that PECD occurring for certain molecular orientations is strongly enhanced and in principle could reach 100\%. In order to verify this prediction experimentally, one would need to fix the orientation of the examined molecule in space. The present work makes a first step towards this goal by studying uniaxially oriented methyloxirane molecules upon O($1s$)-photoionization ($\hbar\omega=550$~eV) using the COLTRIMS-technique \cite{r26} with a specially designed high-resolution (3d focusing for electrons and ions)  spectrometer without any meshes in order to increase the overall particle detection efficiency. The peak of observed photoelectrons was centered at a kinetic energy of about 11.5~eV. Before the nuclei start to rearrange in response to the creation of the O($1s$)-hole, an ultrafast Auger decay takes place, which is finally followed by a Coulomb fragmentation of the doubly-charged ion. {Even though fragments with the mass over charge ratio $m/z$ equal to} 14, 15, 25, 26, 27, 28, 29, 30, 31, and 42 have been observed in the Photoion-Photoion Coincidence (PIPICO) spectra, the present analysis was performed only for two types of molecular breakup with the following fragment combinations:  $C_2 H_2^+ (m/z~26)- COH^+ (m/z~29)$ and $CH_3^+ (m/z~15)-C_2 H_2 O^+ (m/z~42)$. For such a rather large molecule it is not straightforward to relate the measured asymptotic momenta of the ionic fragments to a given molecular axis. Therefore, the molecular orientation at the instant of the photoionization was an optimization parameter in the present electronic structure and dynamics calculations, which were carried out by the Single Center method and code \cite{r27,r28}. Details on the experimental approach and the theory can be found in the Supplemental Materials document \cite{SM}.

\begin{figure}
\includegraphics[scale=0.41]{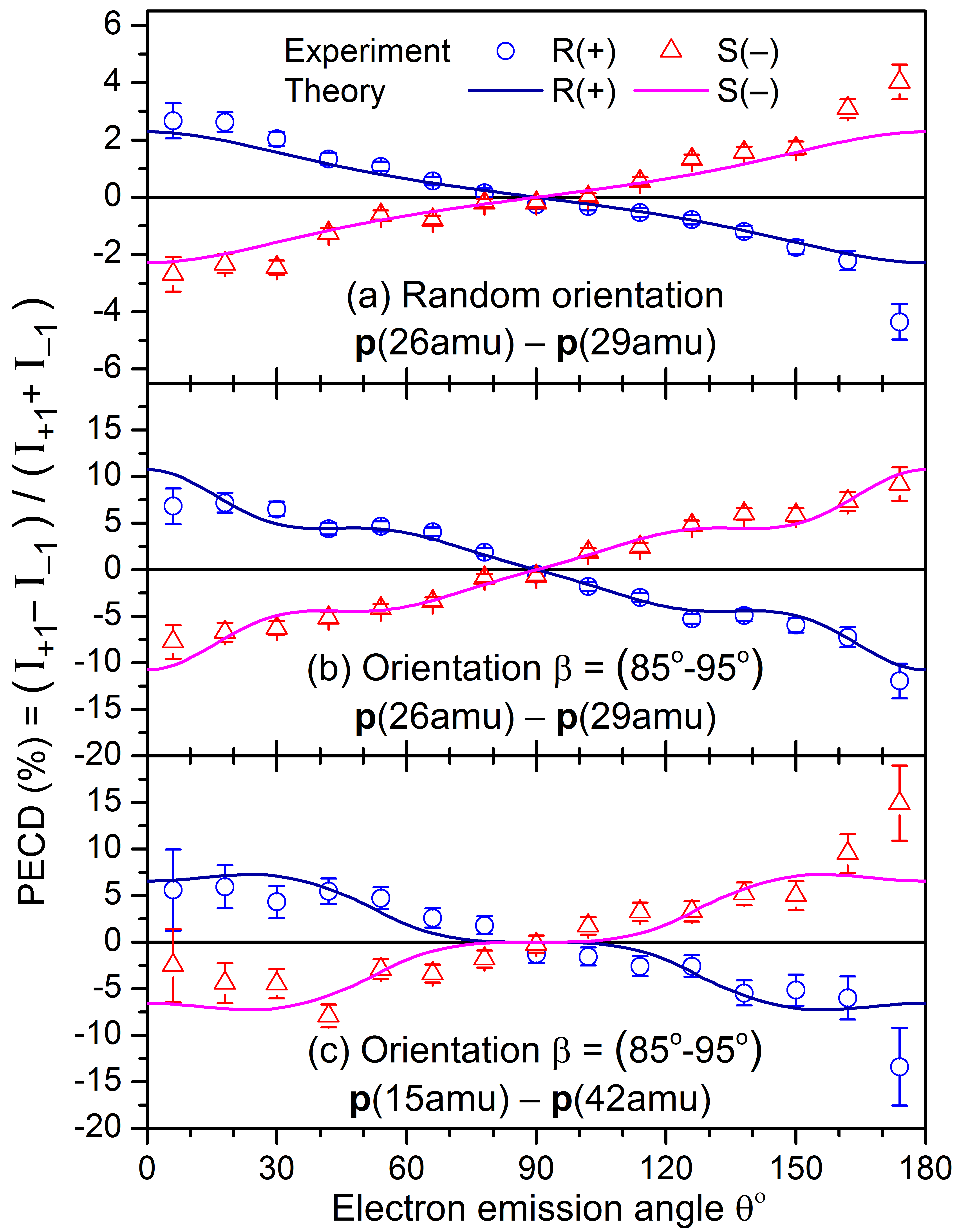}
\caption{(Color online) PECD as a function of the photoelectron emission angle $\theta$, with respect to the photon propagation, measured and computed in the present work for the O 1s-photoionization of the two enantiomers of methyloxirane: (a) for randomly oriented molecules and $p(26amu)-p(29amu)$ fragmentation channel, (b) and (c) for the fragmentation axis being fixed at an angle $\beta =90^\circ$  with respect to the light propagation direction and for the two different fragmentation channels $p(26amu)-p(29amu)$ and $p(15amu)-p(42amu)$.}\label{fig:fig2}
\end{figure}

Fig.~\ref{fig:fig2} compares the measured and computed PECD for randomly oriented molecules (a) and for the two cases where the fragmentation axis of the molecules was fixed-in-space (b and c). The PECD measured for the randomly oriented molecules (note that Fig. \ref{fig:fig2}a comprises only data for the $p(26amu)-p(29amu)$ channel) shows maximum values between 3\% and 4\%, whereas the PECD obtained after fixing the fragmentation axes in space parallel to the polarization plane (the angle between the fragmentation axis and the light propagation axis was between 85 and 95$^\circ$) shows much higher asymmetries (Figs. \ref{fig:fig2}b and \ref{fig:fig2}c). In particular, applying such a restriction to the fragmentation channel $p(26amu)-p(29amu)$ gives experimental asymmetry values up to 12\%, whereas theoretical curves lead to the maximum asymmetry of 10\%. Similar enhancement is observed for the fragmentation channel $p(15amu)-p(42amu)$. We also note that the generally expected change of sign of PECD with respect to the interchange of the enantiomers (R(+) and S(--) denote the two different enantiomers) is { clearly observed for both randomly oriented and fixed-in-space molecules.}

\begin{figure}
\includegraphics[scale=0.44]{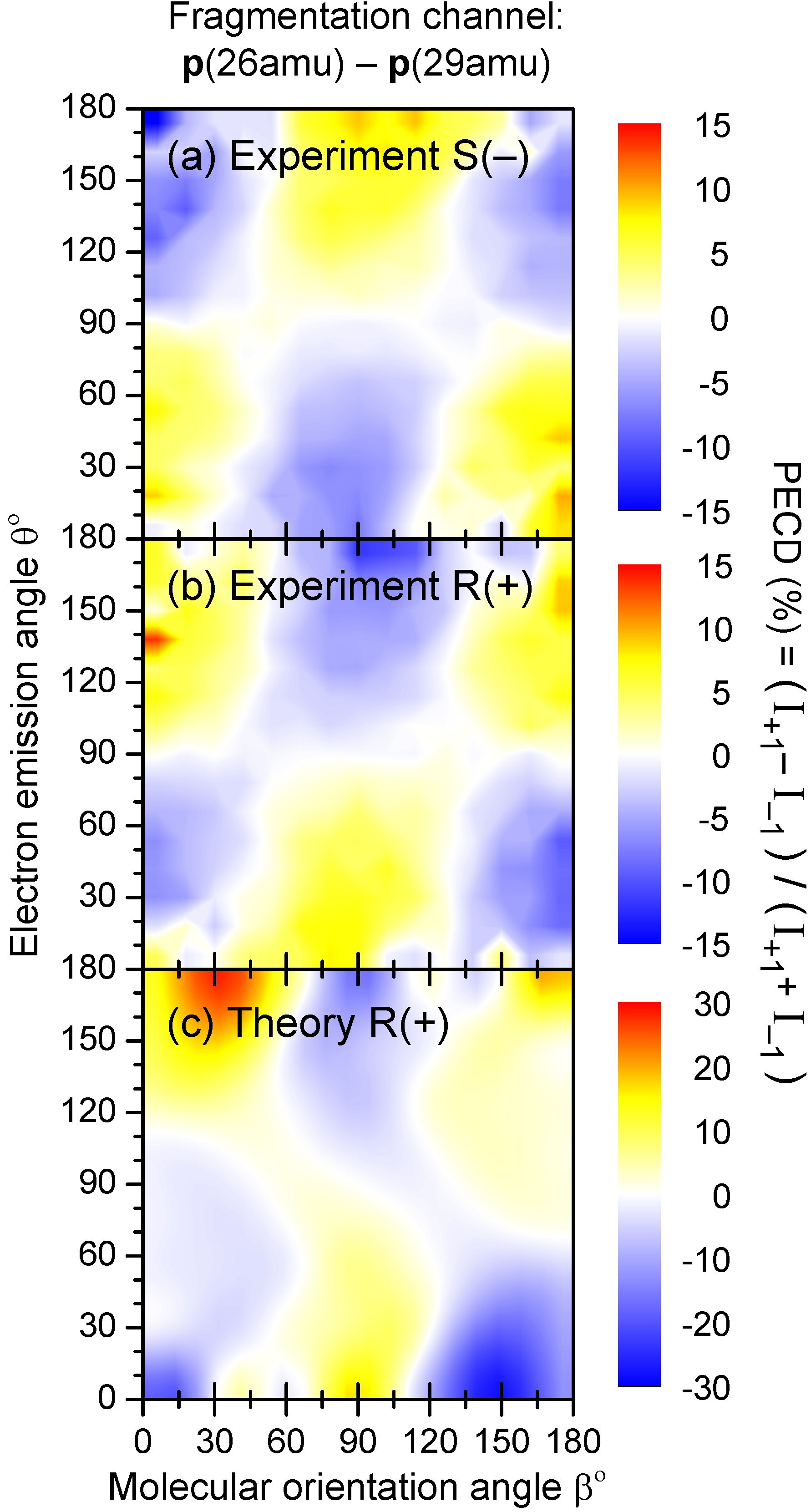}
\caption{(Color online) PECD as a function of the photoelectron emission angle $\theta$ and the molecular orientation angle $\beta$ after O 1s-photoionization and subsequent dissociation of methyloxirane into $p(26amu)$ and $p(29amu)$ fragments: (a) measurements for the S(--) enantiomer; (b) measurements for the R(+) enantiomer; (c) calculations for the R(+) enantiomer. Note, that PECD computed for the S(--) enantiomer (not shown here) has an opposite sign. For the sake of brevity, orientation of the molecular fragmentation axis in space is shortly referred to throughout as the molecular orientation.}\label{fig:fig3}
\end{figure}

\begin{figure}
\includegraphics[scale=0.44]{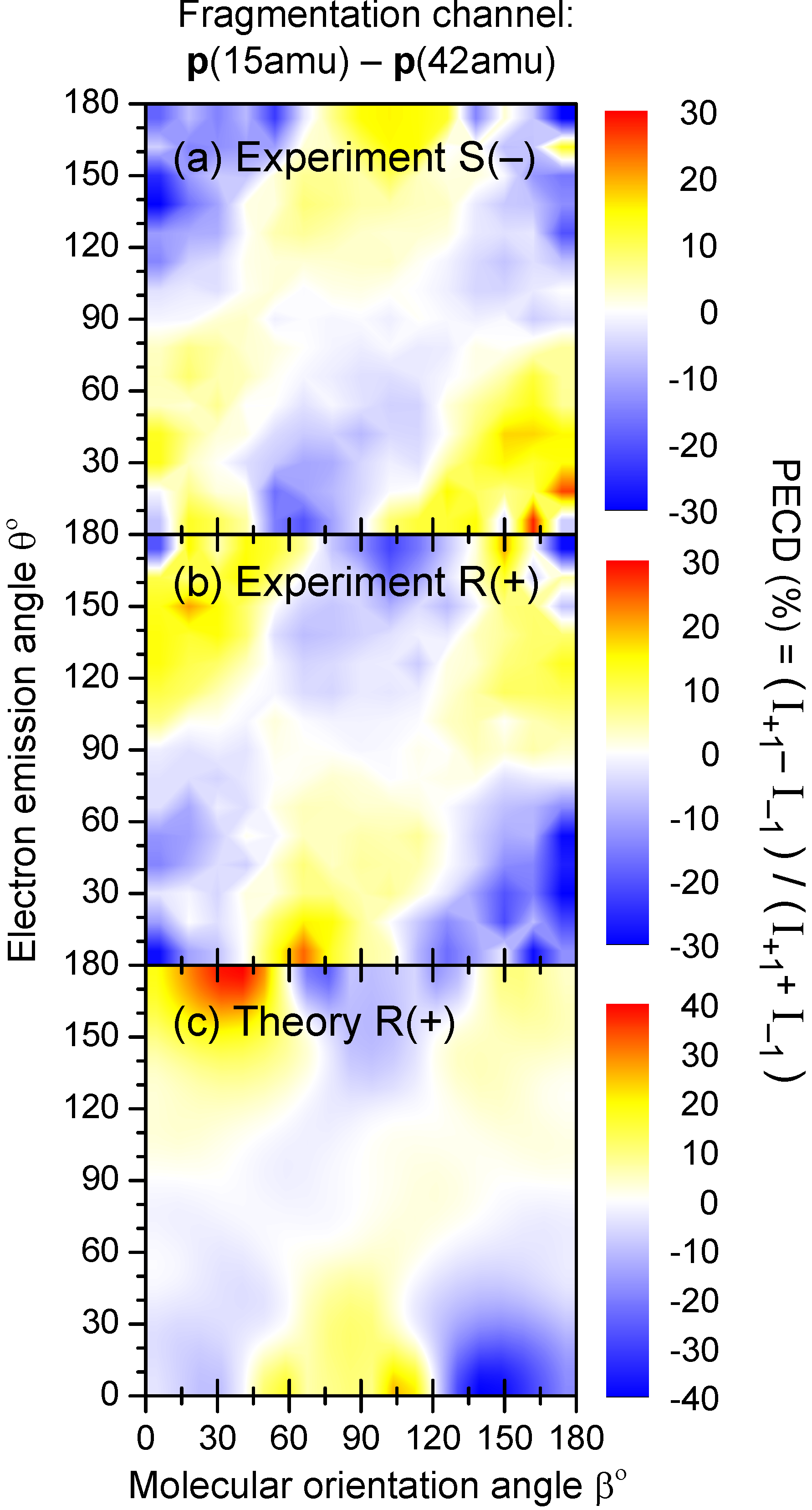}
\caption{(Color online) The same as in Fig.~\ref{fig:fig3} but for the $p(15amu)-p(42amu)$ fragmentation channel.}\label{fig:fig4}
\end{figure}

A more detailed view on the PECD is given in Figs. \ref{fig:fig3} and \ref{fig:fig4}. These figures depict PECDs obtained for the two fragmentation channels as functions of the photoelectron emission angle  $\theta$ and of the molecular orientation angle $\beta$  (angle between the fragmentation axis and the light propagation). One can see, as well, that the sign of the PECD changes when the enantiomers are swapped (cf. Figs. \ref{fig:fig3}a with \ref{fig:fig3}b and Figs. \ref{fig:fig4}a with \ref{fig:fig4}b). This latter finding confirms that the observed asymmetry has a chiral origin. Moreover, the measured two-dimensional PECDs confirm the following analytically derived symmetry property: $PECD(\pi-\theta;\pi-\beta)=-PECD(\theta;\beta)$. Since we observe different signs of PECD upon switching between the forward ($\theta=0^\circ$) and the backward ($\theta=180^\circ$) photoemission directions, the symmetry rule results in similar signs for the molecular axis oriented along the light propagation ($\beta=0^\circ$, forward) and in the reversed direction ($\beta=180^\circ$, backward). For a given enantiomer, the PECD reverses its sign when the molecular orientation changes from being parallel to the light propagation axis ($\beta=0/180^\circ$) to the case when it is orthogonal to the light propagation ($\beta=90^\circ$). Therefore, integration over all orientations results in a considerably smaller effect.

Experimentally, we find an asymmetry of up to 15\% for the $p(26amu)-p(29amu)$ breakup and even higher asymmetries up to 30\% in the case where ($\theta=0^\circ$; $\beta=160^\circ$) for the fragmentation channel $p(15amu)-p(42amu)$. The higher asymmetry observed  at some  molecular orientations for the latter fragmentation channel can be explained by the fact that it corresponds to a single bond breaking (loss of methyl group), whereas the former requires the breaking of two bounds. As a consequence, the analysis of the coincident data allows for a more accurate determination of the molecular orientation for the $p(15amu)-p(42amu)$ breakup, whereas in the case of the $p(26amu)-p(29amu)$ channel, an additional averaging over orientations can be present. The computed PECDs (Figs. \ref{fig:fig3}c and \ref{fig:fig4}c) show a good overall agreement with the experimental data: Both have similar signs of asymmetries, but the theoretical values are somewhat overestimated. For the fragmentation channel $p(26amu)-p(29amu)$, the calculations show asymmetries up to 30\%, whereas PECD of about 35\% is computed for the fragmentation channel $p(15amu)-p(42amu)$. Integration of the signals $I_{\pm1}$ over all angles $\beta$  gives the much smaller PECD observed for randomly oriented samples (Fig. \ref{fig:fig2}a). {Finally, the dichroic parameter $b_1$ computed and measured for R(+) enantiomer is very small and equal to 1.35\% and $1.56$\%$\pm 0.25$\%, respectively.} These results support the intuitive prediction that selecting a particular 3D orientation, rather than averaging over all orientations, enables to remove any cancelation that occurs due to compensation of the PECD for different molecular orientations.

\begin{figure}
\includegraphics[scale=0.40]{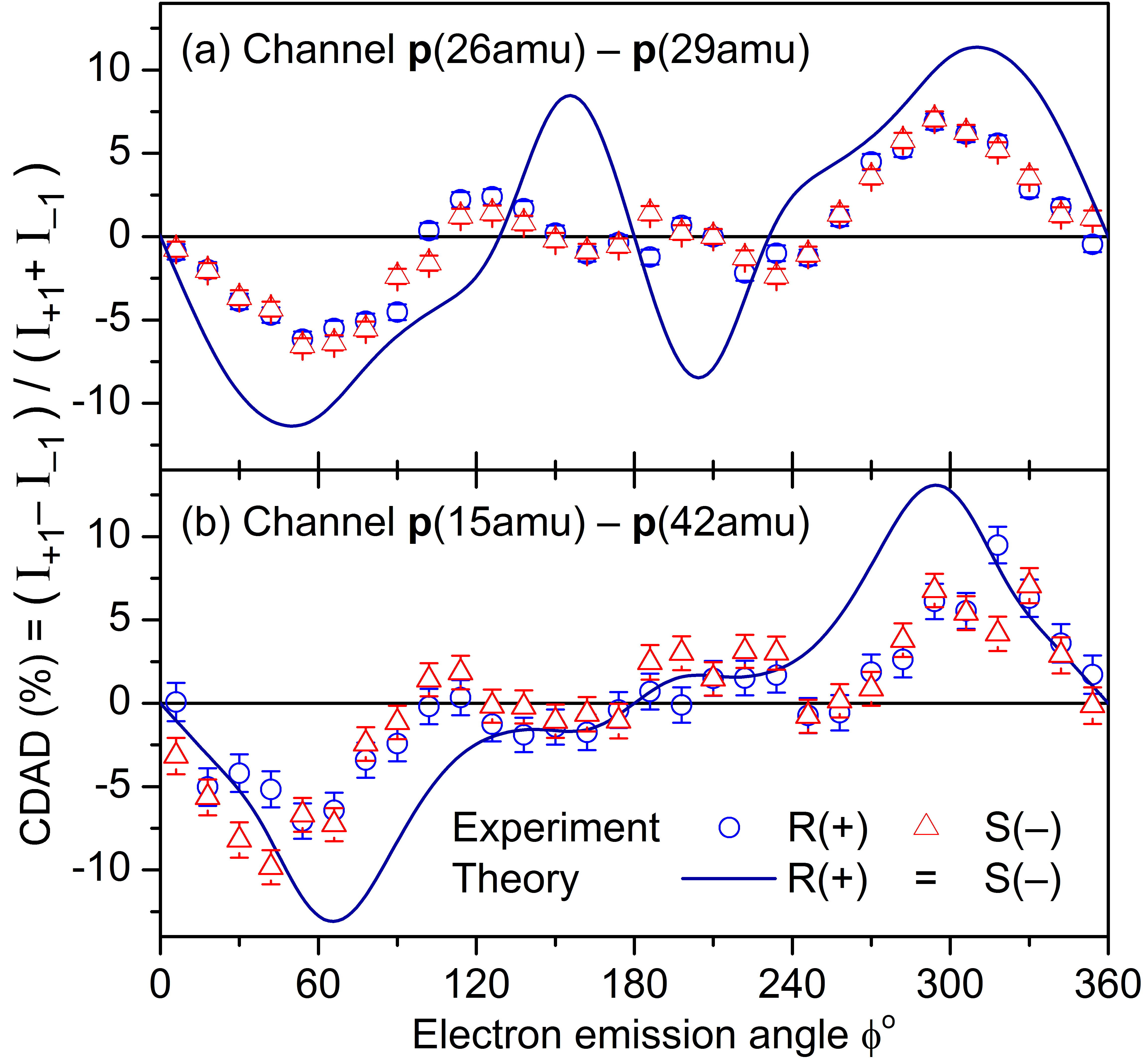}
\caption{(Color online) CDAD as function of the azimuthal photoelectron emission angle $\phi$ in the polarization plane, measured and computed for the O 1s-photoionization of two enantiomers of methyloxirane and the two fragmentation channels: (a) $p(26amu)-p(29amu)$ and (b) $p(15amu)-p(42amu)$. Note, that CDAD computed for R(+) and S(--) enantiomers are equivalent. Electrons are selected in the range of $0^\circ <\theta< 180^\circ$ and ions in the range of $85^\circ <\beta< 95^\circ$. }\label{fig:fig5}
\end{figure}

As PECD depends only on the angle $\theta$ with respect to the light propagation axis, we have so far averaged the electron distribution over the azimuthal angle  $\phi$. By selecting molecular orientations to be perpendicular to the light propagation direction ($\beta = 90^\circ$), the CDAD as a function of the azimuthal emission angle  $\phi$ in the polarization plane can be extracted from the experimental coincident data. The measured CDADs are depicted in Fig. \ref{fig:fig5}. Unlike PECD, which have different signs for the two enantiomers, the CDAD is enantiomer insensitive, having equal trends for R(+) and S(--) methyloxirane. Accordingly, the CDAD has to vanish for randomly oriented chiral molecules, similarly to the case of randomly oriented achiral molecules. Finally, the computed CDADs reproduce the trends of the experimental asymmetry, although the theory slightly overestimates its magnitude (Fig. \ref{fig:fig5}). {We notice that CDAD and PECD were obtained by two very different data treatments (see Supplemental Materials \cite{SM} for details). This can be a reason for the larger disagreement between the theory and experiment for CDAD, since building a new coordinate system for the data analysis could result in larger uncertainties of the emission and orientation angles determination. The observed discrepancy therefore exceeds the purely statistical error bars shown in Fig.~\ref{fig:fig5}.}

{ In conclusion, all previous studies of PECD in the gas phase were performed for randomly oriented chiral molecules. In those studies, PECD was discussed in terms of laboratory frame angular distribution and described as a forward/backward asymmetry in the photoelectron emission which survives after averaging over all molecular orientations. Our theoretical predictions illustrate that fixing three-dimensional orientation of a target in space may in principle result in a 100\% effect, as it is known for CDAD. Using coincident detection technique we provide the first experimental proof for those expectations and demonstrate that chiral asymmetry for O(1s)-photoionization of methyloxirane can be significantly enhanced already by fixing one molecular fragmentation axis. Providing larger asymmetries makes PECD of oriented chiral molecules a more sensitive tool for the enantiomeric excess determination. The present analysis supports the transparent picture of the photoelectron scattering on the molecular potential being at the heart of the PECD. By interrelating the fundamental PECD and CDAD phenomena with the molecular frame photoelectron angular distribution we pave the way for a detailed understanding of the origin of this fundamental photo physical effect.}

\begin{acknowledgements}
The work was supported by the State of Hesse Initiative for the Development of Scientific and Economic Excellence (LOEWE) within the focus-project Electron Dynamics of Chiral Systems (ELCH). Financial support by the Deutsche Forschungsgemeinschaft (DFG Project No. DE~2366/1-1) is gratefully acknowledged. We thank the staff of SOLEIL for running the facility and providing beamtimes under project 20140056 and 20141178 and especially SEXTANTS beamline for their excellent support. M. S. gratefully acknowledges support by the Adolf Messer foundation. H.F. and K.U. are grateful for support from the X-ray Free Electron Laser Priority Strategy Program of MEXT and from IMRAM, Tohoku University. We thank Thomas Baumert for suggesting to us the nut/thread analogy, which was originally discussed by Ivan Powis in Ref.~\cite{r16}. We thank T. Daniel Crawford for providing us with the optimized geometry of methyloxirane.
\end{acknowledgements}


\begin{thebibliography}{99}
\bibitem{r07}	
The asymmetry therefore exists without the need to invoke weaker higher order radiation-molecule interaction terms.
\end{thebibliography}
\end{document}

\title{Supplemental Material\\ ~\\
Observation of enhanced chiral asymmetries in the inner-shell photoionization of uniaxially oriented methyloxirane enantiomers \\ ~}

\author{Maurice Tia$^1$, Martin Pitzer$^{1,2}$, Gregor Kastirke$^1$, Janine Gatzke$^1$,  Hong-Keun Kim$^1$,\\ Florian Trinter$^1$,  Jonas Rist$^1$,
Alexander Hartung$^1$, Daniel Trabert$^1$,\\ Juliane Siebert$^1$, Kevin Henrichs$^1$, Jasper Becht$^1$,
Stefan Zeller$^1$, Helena Gassert$^1$, Florian Wiegandt$^1$, Robert Wallauer$^1$, Andreas Kuhlins$^1$,
Carl Schober$^1$, Tobias Bauer$^1$, Natascha Wechselberger$^1$, Phillip Burzynski$^1$, Jonathan Neff$^1$,
Miriam Weller$^1$, \\ Daniel Metz$^1$, Max Kircher$^1$, Markus Waitz$^1$, Joshua B. Williams$^3$, Lothar Schmidt$^1$,\\
Anne D. M\"{u}ller$^2$, Andre Knie$^2$, Andreas Hans$^2$, Ltaief Ben Ltaief$^{\,2}$, Arno Ehresmann$^2$, Robert Berger$^4$,
Hironobu Fukuzawa$^5$, Kiyoshi Ueda$^5$, Horst Schmidt-B\"{o}cking$^1$,\\ Reinhard D\"{o}rner$^1$, Till Jahnke$^1$,  Philipp V. Demekhin$^2$, and Markus Sch\"{o}ffler$^1$\\~}

\affiliation{$^1$Institut f\"{u}r Kernphysik, Universit\"{a}t Frankfurt, Max-von-Laue-Stra{\ss}e 1, 60438 Frankfurt am Main, Germany\\
$^2$Institut f\"{u}r Physik und CINSaT, Universit\"{a}t Kassel, Heinrich-Plett-Str. 40, 34132 Kassel, Germany\\
$^3$Department of Physics, University of Nevada, Reno, Nevada 89557, USA\\
$^4$Theoretical Chemistry, Department of Chemistry, Universit\"{a}t Marburg, Hans-Meerwein-Stra{\ss}e, 35032 Marburg\\
$^5$Institute of Multidisciplinary Research for Advanced Materials, Tohoku University, Sendai 980-8577, Japan}

\maketitle

In this supplemental material document, a few essential details of the present experiment and analysis of data, as well as computational methods and theoretical analysis are discussed.

\section*{Experimental Method}

The experiment was performed using the well-established COLTRIMS (Cold Target Recoil Ion Momentum Spectroscopy) technique~\cite{COLTRIMS} during the 8-bunch mode (pulsed-operation) at the Synchrotron SOLEIL (Saint-Aubin, France). A supersonic expansion of enantiopure methyloxirane molecules (Sigma Aldrich, 99\% purity) is skimmed to form a molecular beam which crosses the synchrotron radiation provided by the variable polarization undulator-based beamline SEXTANTS for alternate light helicities (the polarization was switched every 2 hours). The charged fragments resulting from the Coulomb explosion and the photoelectrons are accelerated in opposite directions perpendicular to the molecular and photon beams by a static electric field (124 V/cm) onto two position and time sensitive multichannel plate detectors (MCPs) with delay line multi-hit readout and an acceptance angle of $4\pi$  for the  photoelectrons and ions. From the position of impact onto the detectors, the known distance between the ionization region and the MCPs, and the time-of-flight, the particle trajectories can be calculated and the momentum vector of each particle determined.

\begin{table}[b]
\caption{Number of events for each fragmentation channel, enantiomer, and polarization of light.}
\begin{ruledtabular}
\begin{tabular}{lrr}
Enantiomer&	Right Circular Polarized Light &	Left Circular Polarized Light\\ \hline
Channel $p(26amu)-p(29amu)$\\
S(--)~~Methyloxirane	&2140188	&2233919\\
R(+) Methyloxirane	&2256934	&1916846\\ \hline
Channel $p(15amu)-p(42amu)$\\
S(--)~~Methyloxirane &	299125	&310940\\
R(+) Methyloxirane	&   309512	&264301
\end{tabular} \label{tab:stat}
\end{ruledtabular}
\end{table}

In our measurement, we have observed and gated our data on the photoelectron peak at 11.5 eV of kinetic energy. As mentioned above, these slow electrons are detected with a solid angle of $4\pi$.  The solid angle of detection of the fast Auger electrons, on the other hand, is very small (acceptance angle of approx. 1\% - 2\%). In the cases where an Auger electron was detected, we are able to discriminate this electron from the photoelectron due to its high energy (i.e., by applying the aforementioned gate on the electron energy in our off-line analysis). In the complete photoion-photoion coincidence spectra (PIPICO), we have also observed: 14 amu ($CH_2^+$), 15 amu ($CH_3^+$), 25 amu ($C_2H^+$), 26 amu ($C_2H_2^+$), 27 amu ($C_2H_3^+$), 28 amu ($CO^+$), 29 amu ($COH^+$), 30 amu ($COH_2^+$), 31 amu ($COH_3^+$), and 42 amu ($C_2H_2O^+$) fragments. Only photoelectrons detected in coincidence with two positive ions $C_2H_2^+ (m/z~26)$ and $COH^+(m/z~29)$  (referred to as the $p(26amu)-p(29amu)$ fragmentation channel) or with $CH^+_3 (m/z~15)$ and $C_2H_2O^+(m/z~42)$  (referred to as the $p(15amu)-p(42amu)$ fragmentation channel) were considered in the present analysis. The numbers of events for the two studied fragmentation channels are listed in Table~\ref{tab:stat}. Approximately $2\cdot 10^6$  events were obtained for the fragmentation channel $p(26amu)-p(29amu)$ and $3\cdot10^5$ events for the fragmentation channel $p(15amu)-p(42amu)$ for a given enantiomer and a given helicity of the light, which determines the statistical error bars in Figs. 2 and 5 of the main manuscript. The error bars arise indeed from the image counting statistics which are modeled with the Poisson distribution. The propagation of the errors through the normalized difference gives the observed error bars.

The PECD in Figs.~3 and 4 of the main manuscript is obtained by building a 2D histogram in which the vertical axis depicts the electron emission angle $\theta$ as a function of the molecular orientation $\beta$. The PECD in Figs. 2b and 2c of the main manuscript is  obtained by a projection of this histogram in the range ($85^\circ-95^\circ$). For the CDAD in  Fig. 5 of the main manuscript, a new right-handed coordinate system was built with one axis being the light propagation axis, the second is the fragmentation axis $p_1-p_2$, and the third one was set orthogonal to the two others. In this coordinate system, we extracted the angular distribution of photoelectrons in the polarization plane and determined the CDAD.

\section*{Theoretical Method}

\begin{table}[b]
\caption{Equilibrium internuclear geometry (in bohr) of the ground electronic state of S(--) methyl\-oxirane optimized by the CCSD(T)/cc-pVTZ method and basis in Ref.~\cite{GEOM}.}
\begin{ruledtabular}
\begin{tabular}{llrrr}
Atom&Charge& $x~~~~~~~$  &$y~~~~~~~$& $z~~~~~~~$\\ \hline
     O   &  8   &  -1.53014963  &    1.38635664   &   0.26811184\\
     C   &  6   &  -1.80055939  &   -1.30519800   &   0.21822630\\
     C   &  6   &   0.39536858  &   -0.08152896   &  -0.92658199\\
     C   &  6   &   2.92322848  &   -0.09078644   &   0.34281369\\
     H   &  1   &   3.93336581  &    1.65888792   &  -0.04159667\\
     H   &  1   &   4.06602933  &   -1.66226789   &  -0.33623256\\
     H   &  1   &   2.68587329  &   -0.27205357   &   2.37640354\\
     H   &  1   &   0.41409323  &    0.09847784   &  -2.96352555\\
     H   &  1   &  -3.31116837  &   -2.00291144   &  -0.96478056\\
     H   &  1   &  -1.57862960  &   -2.23011356   &   2.02704968
\end{tabular} \label{tab:geom}
\end{ruledtabular}
\end{table}

In order to describe the O 1s-photoionization of the methyloxirane enantiomers, we applied the ab initio theoretical approach developed in our previous angle-resolved studies of the core-excited diatomic~\cite{dia1,dia2} and polyatomic molecules \cite{poly1,poly2,poly3}. The electronic structure and dynamics calculations were carried out by the Single Center method and code \cite{SC1,SC2} which provides an accurate description of the partial photoelectron continuum waves in molecules. The photoionization transition amplitudes were computed in the Relaxed-Core Hartree-Fock approximation including monopole relaxation of the molecular orbitals in the field of the core-vacancy. The calculations were performed at the equilibrium \cite{GEOM} internuclear geometry of the ground electronic state of methyloxirane (listed in Tab.~\ref{tab:geom}).

\begin{figure}
\includegraphics[scale=0.48]{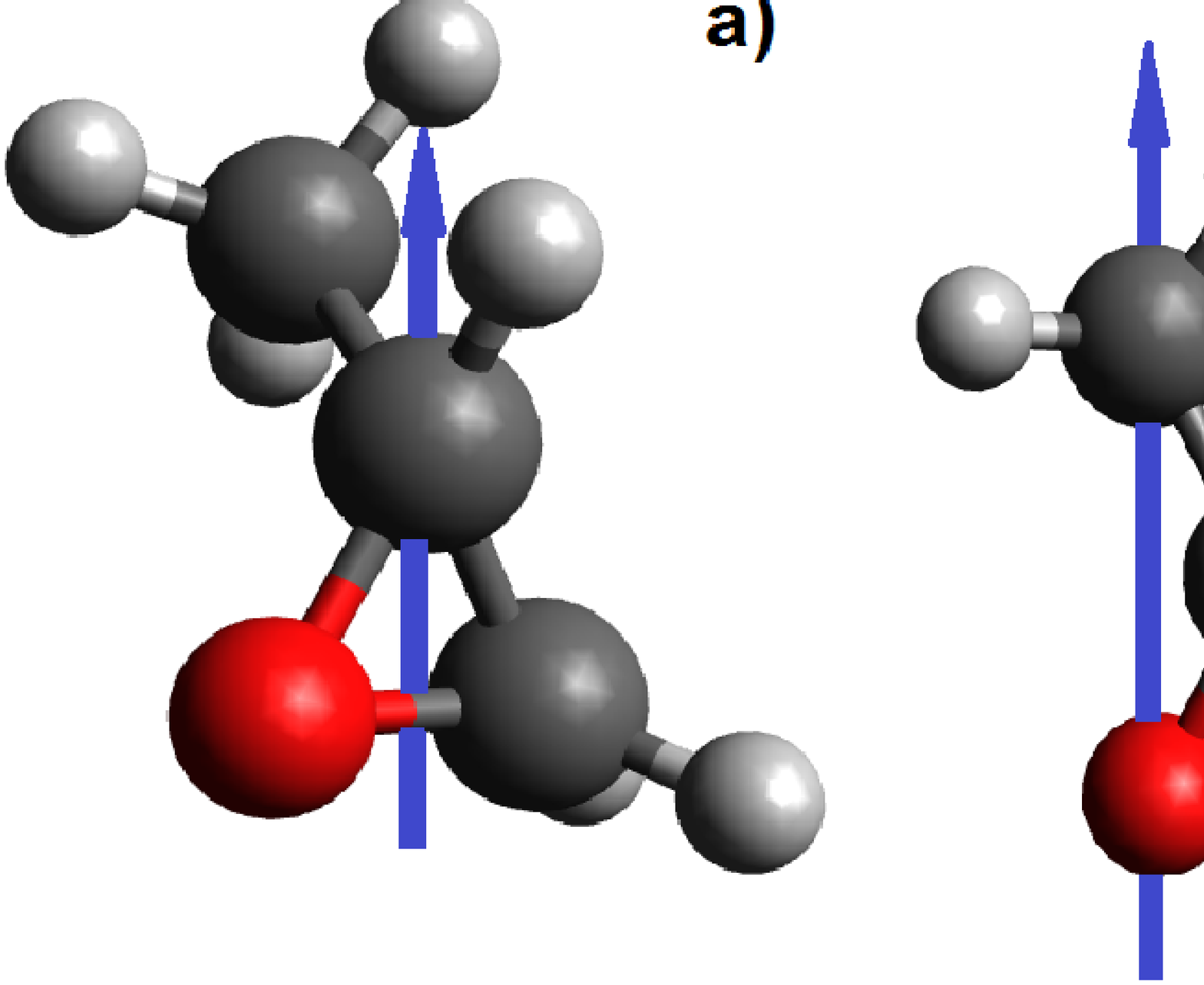}
\caption{The theoretically found molecular orientation axis at the instant of the O 1s-photoionization of the R(+) methyloxirane enabling to reproduce the experimental PECD and CDAD in  Figs. 2 -- 5 of the main manuscript. Panel a): for the $p(26amu)-p(29amu)$ channel, the axis practically coincides with the median of the OCC triangle with the CH group at the top. Panel b): for the $p(15amu)-p(42amu)$ channel, it connects the oxygen atom with the carbon of the methyl group.} \label{fig_axes}
\end{figure}

Calculations of the partial O 1s-photoionization transition amplitudes were performed for different orientations of the molecular frame (MF) with respect to the laboratory frame (LF, $Z_{LF}$ is defined by the direction of the propagation of the exciting-radiation). These orientations are defined by the two Euler angles $\alpha$ and $\beta$, whereas the third angle $\gamma$ (which defines a rotation around the laboratory $Z_{LF}$ axis) is irrelevant for the present experimental geometry. The orientation intervals $\alpha \in [0,2\pi)$ and $\beta\in [0,\pi]$ were  covered with the steps of a $\Delta\alpha=\Delta\beta=0.025\pi$. The theoretical photoelectron circular dichroism (PECD) shown in Figs.~3 -- 4 of the main manuscript was obtained by numerical integration over the orientation angle $\alpha$, whereas PECD in Fig.~2 and circular dichroism in the angular distribution (CDAD) in Fig.~5 required additional integration over the angle $\beta$.

As not all of the particles are detected in the present coincident experiment and as the fragmentation process might be rather complicated, it is difficult to relate the experimental axes determined from the asymptotic momenta of the two detected ionic fragments to a given molecular axis. Therefore, we used the molecular orientation axis at the instant of the photoionization process as a free parameter in the calculations, and searched for the best fit between the computed and measured PECD and CDAD. The visual correspondence between the present theory and experiment was used as the convergence criterium. The molecular orientation axes obtained in the present calculations are schematically illustrated in Fig.~\ref{fig_axes} for both fragmentation channels. In order to produce the theoretical results shown in Figs.~3 -- 5, these optimized molecular orientations in each case were set to coincide with  $Z_{LF}$ axis, i.e., $\alpha=\beta=\gamma=0$. For an arbitrary molecular orientation, the photoionization transition amplitudes computed at this initial orientation were transformed to the LF by standard Wigner's rotational matrices. Finally, the theoretical results were convolved with the Gaussian function of $5^\circ$ FWHM in order to account for the experimental uncertainty for the angle determination.

\section*{Author contributions}
The experiment was conceived by M.S. and R.D. The experiment was prepared and carried out by M.T., M.P., G.K., J.G., H-K. K., F.T., J.R., A.H., D.T., J.S., K.H., J.B., S.Z., H.G., F.W., R.W., A.Ku., C.S., T.B., N.W., P.B., J.N., M.W., D.M., M.K., M.W., J.B.W., A.Kn., A.H., L.B.L., H.F., K.U. and R.B. Data analysis was performed by M.T., M.P. and M.S. Theoretical calculations were performed by P.V.D and A.D.M. The manuscript was written by M.T., T.J., P.V.D., R.D. and M.S. All authors discussed the results and commented on the manuscript.

\makeatletter
\renewcommand*{\@biblabel}[1]{[S#1]}
\makeatother